\documentclass[amsmath,amssymb,preprint]{revtex4}

\begin{document}

\setlength{\unitlength}{1mm}
\textwidth 15.0 true cm
\textheight 22.0 true cm
\headheight 0 cm
\headsep 0 cm
\topmargin 0.4 true in
\oddsidemargin 0.25 true in
\thispagestyle{empty}

\def\beq{\begin{eqnarray}}
\def\eeq{\end{eqnarray}}

\def\pr{\partial}
\def\Mpl{M_{\rm Pl}}
\def\Rbr{R}
\def\Rbu{{\cal R}}
\def\M{M_*}
\def\gn{G_N}
\def\tgn{{\tilde G}_N}
\def\goesto{\rightarrow}
\def\eps{\epsilon}
\def\x{\bf x}

\title{Non-Gaussianity of Large-Scale Cosmological Perturbations}

\author{Andrei Gruzinov}

\address{Center for Cosmology and Particle Physics, Department of Physics, New York University, NY 10003}

\date{July 31, 2003}

\setcounter{footnote}{0} \setcounter{page}{1}
\setcounter{section}{0} \setcounter{subsection}{0}
\setcounter{subsubsection}{0}

\begin{abstract}

 Perturbations in a universe with matter and cosmological constant are computed in terms of the primordial scalar perturbation up to second order. The calculation is easy -- a generalization of the spherical dust solution. Our expression might allow accurate measurements of the primordial non-Gaussianity. 
 
\end{abstract}
\maketitle

\section{Introduction}

Inflationary models \cite{inflation} explain the origin of cosmological perturbations, and predict the primordial super-horizon non-Gaussianity which is small in the standard one-field slow-roll case \cite{maldacena}. A simple and natural extension of the standard model, however, predicts a much larger non-Gaussianity \cite{dgz,zal}. To measure the primordial non-Gaussianity one needs to compute the current sub-horizon perturbations in terms of primordial super-horizon perturbations up to second order in the amplitude. This is done here for the universe filled with matter and cosmological constant. The pure matter case has been computed by \cite{gks}; in that paper we also discuss observations of non-Gaussian effects in galaxy surveys.

Neglecting tensor (gravitational) perturbations, we assume that primordial perturbations are pure scalar. Following \cite{maldacena,bst,sb}, we use a gauge invariant variable $\zeta$ that remains constant outside the horizon to characterize the primordial perturbations. On super-horizon scales $\zeta$ is defined as follows: $e^{\zeta }$ is proportional to the local scale factor measured on uniform local Hubble parameter hypersurfaces. 

The degree of non-Gaussianity of primordial perturbations $f_{nl}$ can be defined by assuming that $\zeta$ is of the form \cite{komsper} 
\begin{equation} \label{primng}
\zeta=g-(3/5)f_{nl}g^2,
\end{equation}
where $g$ is Gaussian. 
The standard one-field slow-roll inflation predicts $f_{nl}\lesssim 0.05$ \cite{maldacena,spergel}. Other inflationary models \cite{dgz,zal} predict $|f_{nl}|\gtrsim $ few. The WMAP results are $-58<f_{nl}<134$ at 95\% confidence \cite{komatsu}. The Sloan Digital Sky Survey should provide similar accuracy, and future galaxy surves can be much better \cite{roman}. 

The non-Gaussianity of sub-horizon density perturbations is due to Newtonian evolution, relativistic effects, and primordial non-Gaussianity. As we will see, the ratio of relativistic to primordial contributions is $\sim 1/f_{nl}$; accurate determination of $f_{nl}$ must include relativistic effects.

In \S2 we derive the second-order expression for the density perturbation in terms of $\zeta$. In \S3 we generalize the standard Newtonian calculation of the density skewness.

\section{Second-order density perturbation in terms of $\zeta$.}

\subsection{Results}

Consider flat matter plus cosmological constant universe, with no primordial gravitational waves. We use the comoving synchronous gauge. The first-order metric is
\begin{equation} \label{metric}
ds^2=dt^2-a_0^2(t)(1+2\zeta )(\delta _{ij}+2\tau \zeta _{ij})dx_idx_j.
\end{equation}
Here $a_0(t)$ is the unperturbed scale factor; $\zeta (x)$ is the primordial perturbation, which is a function of $x$ only; $\zeta _{ij}$ indicates spatial derivatives; $\tau (t)$ is the linear growth factor given in the Appendix. For zero cosmological constant, $\Lambda =0$, we have $\tau \propto t^{2/3}$ -- the familiar linear growth of perturbations in the matter dominated universe. 

The density perturbation to second order is
\begin{equation} \label{delta}
\delta =-\tau \zeta _{ii}+{\tau ^2\over 2}\left( (1+{\cal C})\zeta _{ii}^2+(1-{\cal C})\zeta _{ij}^2\right) +{\tau \over 2}\left( 4\zeta \zeta _{ii}-\zeta _i^2\right) .
\end{equation}
Here ${\cal C}(t)$ is a dimensionless numerical factor given in the Appendix. For $\Lambda =0$, ${\cal C}=3/7$ and we recover \cite{gks}.

The first two terms in (\ref{delta}) represent the second-order Newtonian perturbation in the synchronous coordinate system \cite{peebles,matarrese}. The third term represents the relativistic effects. The first-order metric can be derived along standard lines \cite{mukh}. The second order density perturbation is obtained in the next subsection. The applications of this expression are discussed in \S3 and \S4 .

\subsection{Derivation}

We calculate the second order density perturbation under spherical symmetry. Then we simply generalize to the non-spherical case.

Consider the well known exact solution of the Einstein equations -- the spherically symmetrical dust \cite{dingle,peebles} with cosmological constant. The metric is
\begin{equation} \label{metrics}
ds^2=dt^2-{(ax)'^2\over 1-x^2/R^2}dx^2-a^2x^2(d\theta ^2+\sin ^2\theta d\phi ^2).
\end{equation}
The matter density is 
\begin{equation} \label{density}
\rho = {3(x^3F)'\over 8\pi G(x^3a^3)'}.
\end{equation}
Here $R(x)$ and $F(x)$ are arbitrary functions of $x$ but not of $t$, while $a(x,t)$ and $\rho (x,t)$ are functions of both $x$ and $t$, and $a$ satisfies the Friedmann equation
\begin{equation} \label{scale}
a\dot{a} ^2+a/R^2=F+\Lambda a^3/3.
\end{equation}
Dot and prime are time and $x$ derivatives.

We can choose 
\begin{equation} \label{F}
F\propto e^{3\zeta }, 
\end{equation}
then at $t\rightarrow 0$ equation (\ref{scale}) gives $a\propto e^{\zeta }t^{2/3}$. If we also choose 
\begin{equation} \label{R}
R^{-2}=-{2\over x}\zeta '-\zeta '^2, 
\end{equation}
equation (\ref{metrics}) gives the early time (superhorizon) metric of the form
\begin{equation} \label{metricz}
ds^2=dt^2-a_0^2(t)e^{2\zeta }d{\bf x}^2,
\end{equation}
where $a_0(t)\propto t^{2/3}$, $t\rightarrow 0$, is the unperturbed scale factor. We see that what we have called $\zeta$ is indeed the standard gauge invariant quantity that inflation theories like \cite{maldacena,dgz} claim to predict to second or arbitrary order.

Having expressed the two free parameters of the spherical model, $R(x)$ and $F(x)$, in terms of $\zeta (x)$, we can now calculate the spherical density perturbations in terms of $\zeta$ to second order. The dimensionless density perturbation $\delta$ is defined by
\begin{equation} \label{deltadef}
\rho \equiv (1+\delta )\rho _0,
\end{equation}
$\rho _0$ is the unperturbed matter density. It is convenient to define $\delta _a$ by
\begin{equation} \label{deltaadef}
a\equiv e^{\zeta }a_0(t)(1+\delta _a).
\end{equation}
Then equation (\ref{density}) gives
\begin{equation} \label{delta2}
\delta = -3\delta _a-x\delta _a'+6\delta _a^2+4x\delta _a \delta _a'+x^2\delta _a'^2+x^2\zeta '\delta _a',
\end{equation}
and equation (\ref{scale}) gives
\begin{equation} \label{deltaa2}
\delta _a= \delta _{a1}(1-{\cal C}\delta _{a1}-2\zeta +{1\over 2}x\zeta '),
\end{equation}
where 
\begin{equation} \label{deltaa1}
\delta _{a1}= \tau {1\over x}\zeta '.
\end{equation}

Finally we put (\ref{deltaa2}) into (\ref{delta2}) and get the spherically symmetrical density perturbation to second order
\begin{equation} \label{deltas}
\delta =-\tau (\zeta ''+{2\over x}\zeta ')+\tau ^2(\zeta ''^2+{2+2{\cal C} \over x}\zeta '\zeta ''+{3+{\cal C} \over x^2}\zeta '^2)+{\tau \over 2}(4\zeta \zeta ''+{8\over x}\zeta \zeta '-\zeta '^2).
\end{equation}

To get the corresponding expression without the spherical symmetry assumption, we write down the most general expression for a scalar which contains the same powers of $\zeta$, $\tau$ and $\nabla$ as the expression (\ref{deltas}):
\begin{equation} \label{deltag}
\delta =-\tau C_1\zeta _{ii}+\tau ^2(C_2\zeta _{ii}^2+C_3\zeta _{ij}^2)+\tau (C_4\zeta \zeta _{ii}+C_5\zeta _i^2).
\end{equation}
Now assume that $\zeta$ is spherically symmetrical. Then (\ref{deltag}) should reproduce (\ref{deltas}). We have 5 arbitrary coefficients in (\ref{deltag}) in terms of which we should  get the 8 known coefficients of (\ref{deltas}). If we could do it, it would be a check of our spherically symmetrical expression. We can, and the result is given by (\ref{delta}).

\section{The skewness of the density perturbation}

The skewness of the density perturbation $S\equiv <\delta^3>/<\delta ^2>^2$. The computation should include the Jacobian of (2), $J = 1+3\zeta +\tau \zeta_{ii}$. To lowest order,
\begin{equation} \label{skewness}
S = (4+2{\cal C}) -{ \frac{1}{\tau k^2}} {36\over 5} (f_{nl}+{35\over 24})
\end {equation}
Here $k^2\equiv <\zeta_{ii}^2>/<\zeta_i^2>$. For $\Lambda =0$, neglecting primordial and relativistic contributions of the second term, we obtain the Newtonian result \cite{peebles}, $4+2{\cal C}=34/7$. 

The actual perturbation spectrum is scale-free, and our formula is not directly applicable. However it may serve to compare primordial and relativistic contributions to the sub-horizon non-Gaussianity.

\section{Conclusions}

We have expressed the large-scale cosmological perturbations in terms of the primordial scalar perturbations up to second order. Our expression may allow to measure primordial non-Gaussianities using galaxy surveys \cite{roman}.

I thank Raisa Karasik and Roman Scoccimarro for useful discussions, and P.J.E. Peebles for telling me about the spherical perturbation case. This work was supported by the David and Lucile  Packard Foundation. 

\begin{appendix}

\section{ $\tau (t)$ and ${\cal C}(t)$}

The linear growth rate in the synchronous comoving gauge coincides with the Newtonian expression for perturbations in a universe with matter and cosmological constant. We just need to re-derive it by perturbing eq. (\ref{scale}) according to (\ref{deltaadef}). To first order we get
\begin{equation} \label{firstordergrowth}
\dot{a} _0\dot{\delta }_{a1}+4\pi G\rho _0a_0\delta _{a1}={1\over a_0}{\zeta '\over x}.
\end{equation}
We write the solution in the form (\ref{deltaa1}). Then
\begin{equation} \label{growth1}
\tau(t) \equiv D(t)\int _0^t{dt_1\over D(t_1)a_0(t_1)\dot{a_0} (t_1)},
\end{equation}
where $D(t)$ is another growth factor (the homogeneous solution of (\ref{firstordergrowth})):
\begin{equation} \label{growth2}
\ln D(t)= -4\pi G\int ^t{dt_1 \rho _0(t_1)a_0(t_1) \over \dot{a_0} (t_1)}.
\end{equation}
These equation coincide, after simple rearrangements, with the expressions given in \cite{peebles}. On the other hand, the cosmological perturbation theory \cite{mukh} gives the growth factor of perturbations in the comoving synchronous gauge 
\begin{equation} \label{growth11}
\dot {\tau }=a_0^{-3}\int dt a_0,
\end{equation}
which is equal to (\ref{growth1} ).

At early time the universe is matter dominated, $\rho _0 =1/(6\pi Gt^2)$ and $a_0 \propto t^{2/3}$. Then eq.(\ref{growth2}) gives $D=t^{-1}$, and eq.(\ref{growth1}) gives the familiar growth factor $\tau =\eta ^2/10\propto t^{2/3}$, where $\eta$ is the conformal time. At a later time, when the cosmological constant accelerates the expansion of the universe, the growth slows down.

When equation (\ref{scale}) is solved to second order, a non-linear growth factor will appear. The second-order solution can be written in the form (\ref{deltaa2}, \ref{deltaa1}), with a dimensionless coefficient  
\begin{equation} \label{C}
{\cal C}(t)={D(t)\over \tau ^2(t)}\int _0^tdt_1{2\tau +a_0^2\dot{\tau } ^2/2+2a_0\dot{a_0} \dot{D} \tau ^2/D - a_0\dot{a_0} \tau \dot{\tau } \over Da_0\dot{a_0} },
\end{equation}
with all quantities under the integral being at time $t_1$. At early time, ${\cal C}=3/7$ and we recover the zero cosmological constant result \cite{gks}.

\end{appendix}

\end{document}